\documentclass{amsart}

\usepackage{amssymb}

\title{Chirality in an $E_8$ model of elementary particles}
\author{Robert Arnott Wilson} 
\date{18th August 2022; this version 12th October 2022}

\address{School of Mathematical Sciences, Queen Mary University of London}
\email{R.A.Wilson@qmul.ac.uk}

\newcommand{\so}{\mathfrak{so}}
\newcommand{\su}{\mathfrak{su}}
\newcommand{\gu}{\mathfrak{u}}
\newcommand{\spl}{\mathfrak{sl}}
\newcommand{\elie}{\mathfrak{e}}
\newcommand{\glie}{\mathfrak{g}}
\newcommand{\RR}{\mathbb R}
\newcommand{\CC}{\mathbb C}

\newcommand{\OO}{\mathbb O}

\newcommand{\rep}{\mathbf}

\newtheorem{spec}{Speculative remark}
\begin{document}
\begin{abstract}
We show how chirality emerges naturally from an embedding of the standard model of
particle physics into $E_{8(-24)}$. The well-known 
argument 
that there is no
chiral theory of fundamental physics in $E_8$ is avoided by implementing chirality not
as a property of the  
complexified Lorentz group, but as a property of
the complex representations of the real Lorentz group, combined with a real scalar.
This avoids the 
problems of complexification, and
ensures that the model is completely contained in the real Lie group. 
\end{abstract}
\maketitle

\section{Introduction}
There have been many attempts to 
build a model of fundamental physics in some form of $E_8$. The `semi-split' real form $E_{8(-24)}$
is particularly popular \cite{Lisi,Lisi2,Chester,MDW}. A thorny issue that often arises is how to implement
chirality in this context. Indeed, it is often said that this cannot be done, since Distler and Garibaldi 
 \cite{DG} have proved that there is no `chiral' theory of fundamental physics in any real form of $E_8$. But
to understand what this really means,  
we need to understand the assumptions that they make. The first three assumptions (ToE1) are that
the gauge group is connected, compact and commutes with $SL(2,\CC)$. 
The fourth assumption (ToE2) is that all elementary particles have spin $0$, $1/2$ or $1$.
The fifth assumption (ToE3) is that the theory is \emph{chiral} in a particular technical sense
that they explain.

Of these, only (ToE2) is a genuinely physical assumption that we cannot do without.
The \emph{fact} of chirality is also a physical assumption, but the \emph{definition} of chirality is 
a mathematical assumption.
The rest are also assumptions about the mathematical model, and are therefore open to question.
The assumption of commutation with $SL(2,\CC)$ 
is the Coleman--Mandula theorem \cite{Coleman}, 
which applies to a wide class of models, and is therefore reasonable. 
The assumption of connectedness is dubious, since chirality is first and foremost a
property of the disconnected Lie group $O(3,1)$, that is, the isometry group of Minkowski
spacetime, or of one of its eight double covers \cite{Pin}. This viewpoint also calls into question the
definition of chirality, which in \cite{DG} is  
instead a property of the connected
Lie group $SL(2,\CC)\times SL(2,\CC)$. The compactness assumption has also been questioned elsewhere
\cite{MDW}.

It is easy to see that without the compactness assumption the result 
does not hold, as the models in \cite{Chester,MDW}
demonstrate. In this paper, however, we accept both the
compactness and connectedness conditions, and question only the appropriateness of the choice of definition of chirality in \cite{DG}.
There are several definitions of chirality in the physics literature, not all of which are 
mathematically equivalent.
There are two decisions to be made, 
first to decide which Lie group or Lie algebra is to be used in the definition,
and, second, whether the definition can be made in the abstract group or algebra, or whether it requires
consideration of the representations. 

\section{Definitions of chirality}
An excellent in-depth analysis of the mathematical basis for physical concepts of helicity, parity, chirality and so on is
available in \cite{Pin}.  We take from this what we need in order to implement a chiral model in a real form of $E_8$.
We first summarise the most important points. Crucially, it is necessary to have four Weyl spinors, not just two,
in order to describe a chiral theory completely. This is recognised by Distler and Garibaldi \cite{DG}, 
who use the group $SL(2,\CC)\times SL(2,\CC)$ so that each factor has two Weyl spinors, making four in all.
But unfortunately they 
complexify the Lie algebra of $E_8$ to do this, so that the restriction to any real form only has two Weyl spinors.
The argument of \cite{DG} 
certainly implies that
any real $E_8$ model must contain all four Weyl spinors, but does not rule out this possibility.

The necessity for distinguishing two concepts of handedness in physics 
arises ultimately from the fact that the isometry group $O(3,1)$ of Minkowski spacetime has four connected components,
so that two binary invariants are required rather than one. There are three binary invariants available, 
associated with the P (parity), T (time) and PT involutions in the group. 
Roughly speaking, the word `parity' is used in physics to describe a
relationship between the P and T components on one hand, and the identity and PT components on the other,
while the word `chirality' is used to describe a relationship between the PT component and the identity component, and can also be used (with care!)
to relate the P component to the T component.

To make this more precise,
 it is necessary to lift the isometry group to one of its double covers in order to model fermions. 
 Thus the P, T and PT symmetries lift to \emph{operators} $P$, $T$ and $PT$ respectively.
 There are eight such double covers
\cite{Pin},
distinguished by the three independent choices of whether
each of 
$P$, $T$ and $PT$ 
squares to $+1$ or $-1$. In four of these cases $P$ and $T$ commute, and in the other four they anti-commute.
For a physically chiral theory in the usual sense, we need the chirality operator $PT$ to anti-commute with the parity operator $P$,
in order for the parity violation of the weak force to be interpreted as chirality.
All four of the anti-commuting cases lie inside the Dirac algebra 
(defined as the complex Clifford algebra $\CC \ell(3,1)$) and 
act on the Dirac spinors (elements of $\CC^4$). 
Here $P$ is implemented as either $\pm\gamma_0$ or $\pm i\gamma_0$, and $PT$ as either
$\pm\gamma_5$ or $\pm i\gamma_5$. Both $P$ and $T$ swap the two Weyl spinors,
while $PT$ acts on each as a scalar. In all cases, the eigenvalue of $PT$ can be used to define the chirality of the associated Weyl spinor.
Both $P$ and $T$ individually reverse the chirality.

The two cases with $i\gamma_5$ lie in one of the real Clifford algebras $C\ell(3,1)$, with $i\gamma_0$, or 
$C\ell(1,3)$, with $\gamma_0$. These two matrix algebras give rise to Lie algebras $\so(3,3)$ and $\so(5,1)$ respectively,
by the standard process of defining the Lie bracket of two trace zero matrices to be the matrix commutator. 
These two cases are discussed in detail in \cite{MDW,Chester},
in the context of $E_{8(-24)}$, but both have a gauge group which is a non-compact form of $Spin(10)$. The two cases with
$\gamma_5$ do not lie in any real Clifford algebra, and both give rise to a Lie algebra $\so(2,4)$.
It is these last two that are considered in \cite{Pin} to give rise to physically viable chiral models,
and attempts are made to distinguish them by proposing suitable experiments.
In this paper we consider them both in the context of $E_8$, and show that both of them
can be implemented in $E_{8(-24)}$ with a compact gauge group $Spin(10)$, or a subgroup thereof.
For most of the paper we treat them together, without making a choice between them, until Section~\ref{CPT},
where the discussion of charge conjugation allows us to propose that only one of them models CP-violation correctly.

\section{Principles of $E_8$ models}
Many models begin by restricting to $\so(12,4)$ in order to separate the bosonic and fermionic parts of the algebra, and then split into
some real form of $\so(10)+\so(6)$ in order to incorporate the 
$SO(10)$ grand unified theory (GUT) in some way \cite{SU5GUT}.
For example, \cite{Chester} splits the algebra as
\begin{align}
\so(12,4) \rightarrow \so(3,3)+\so(9,1)
\end{align}
 and \cite{MDW} as 
 \begin{align}
 \so(12,4) &  \rightarrow \so(5,1)+\so(7,3)\cr
 & \rightarrow  \so(2) + \so(3,1) + \so(4) + \so(3,3)\cr
 & \rightarrow \so(2) + \so(3,1) + \su(2) + \spl(3,\RR),
 \end{align}
  but both may run into difficulties
caused by having non-compact forms of $\so(10)$. 
It is suggested in \cite{MDW} that this is not really a problem, since there are complex scalars acting on all the spinors,
and these can be used to convert all the boosts into rotations, individually rather than collectively. 
Nevertheless, we shall not consider 
these examples further in this paper.

An alternative is to split as 
\begin{align}
\so(12,4) & \rightarrow \so(10)+\so(2,4)\cr
& \rightarrow \so(6) + \so(4) + \so(1,1) + \so(1,3)\cr
& \rightarrow \gu(3) + \su(2) + \so(1,1) + \so(1,3), 
\end{align}
which 
splits into 
the standard model gauge algebra, a copy of the real numbers and the Lorentz algebra.
In terms of a `complex' definition of a chiral theory closely related to that in \cite{DG}, one could say that
the `chirality' of the model arises from the fact that the (half-)spin representations of $\so(10)$ and $\so(2,4)$ are complex of
dimension $16$ and $4$ respectively. However, we reduce to a `real' definition of a chiral model that relates the 
two direct summands of the weak gauge algebra $\so(4) = \su(2)_L \oplus \su(2)_R$ to the two connected components 
of the group $SO(1,1)$. This is a standard definition in terms of the eigenvalues of the volume elements (pseudoscalars)
of the Clifford algebras $C\ell(0,4)$ and $C\ell(1,1)$. These two volume elements define what are called `weak handedness'
and `Lorentz handedness' respectively in \cite{MDW}.

The approach we take in this paper is to combine 
\begin{align}
\gu(3) + \su(2) & \rightarrow \su(5)\cr
\so(1,1) + \so(1,3) & \rightarrow \so(2,4) = \su(2,2)
\end{align}
so that we can combine an explicit copy of the Georgi--Glashow $SU(5)$ GUT with an approach to
spacetime based on 
twistors.
Here we use the word `twistor' as a shorthand for the Penrose (non-projective)
twistors \cite{twistors,twistorlectures}, defined as Weyl spinors of $\so(2,4)$.
The extra feature of $E_{8(-24)}$ that we use to go beyond these models is an embedding
of the above algebras in a maximal subalgebra of type $A_4+A_4$:
\begin{align}
\su(5) + \su(2,2) \rightarrow \su(5) + \su(2,3)
\end{align}
which we believe provides a useful way to model a unification of classical spacetime with the quantum vacuum.
A main theme of the paper is then to try to 
find a reasonable definition of the `handedness' of the quantum vacuum,
and if possible to relate it to the handedness of Minkowski spacetime, or the weak force, or both.

\section{Explicit calculations}
In order to see these algebra decompositions explicitly, 
we use the notation for $E_{8(-24)}$ developed in \cite{WDM},
as modified in \cite{MDW}.
This notation is based on the octonion division algebra $\OO$, with basis $1,i,j,k,l,il,jl,kl$, and the split octonion algebra $\OO'$,
with basis $U,I,J,K,L,IL,JL,KL$, in which both $1,i,j,k$ and $U,I,J,K$ span quaternion algebras, and $l,L$ act as quaternion conjugation
in the respective cases.
We use $\OO+\OO'$ as a notation for the $16$-dimensional vector representation of $\so(12,4)$, so that subalgebras of $\so(12,4)$ can be
easily specified. For any two elements $a,b$ of $\OO$ there is an element $D_{a,b}$ that acts on that $2$-space, and similarly 
$D_{A,B}$ for $A,B$ in $\OO'$. We also write $D_b=D_{1,b}$ and $D_B=D_{U,B}$. For $a$ in $\OO$ and $B$ in $\OO'$ we have
a similar element $X_{aB}$, with obvious shorthand notation dropping $1$ or $U$.

There is a wide choice of copies of the various groups, and it is not clear which copy will provide the best
illustration of the physical symmetries. 
One possibility is to take $\so(4)$ to act on the
symbols 
$l,I,J,K$, so that it mixes complex numbers on $1,l$ with quaternions on $U,I,J,K$. 
Then we can take $\so(6)$ to act on $i,j,k,il,jl,kl$, so that $\so(2,4)$ acts on $1,U,L,IL,JL,KL$.
The latter can be split into $\so(1,1)$ acting on $U,L$, so with generator $D_L$, and $\so(1,3)$
acting on $1,IL,JL,KL$, so with a basis
\begin{align}
X_{IL}, X_{JL}, X_{KL}, D_{IL,JL}, D_{IL,KL}, D_{JL,KL}.
\end{align}
The rest of $\so(2,4)$ then consists of two Lorentz $4$-vectors:
\begin{align}\label{oddDirac}
&X_1, D_{IL}, D_{JL}, D_{KL};\cr
&X_L, D_{L,IL},D_{L,JL},D_{L,KL}
\end{align}

In other words, the Lie algebra $\so(2,4)$ provides us with a good approximation to a real form of the (complex)
Dirac algebra $\CC \ell(3,1)$. Of course, the Dirac algebra itself is associative, so cannot be embedded directly into a Lie algebra,
so that some technical modification is certainly required.
The usual (Clifford algebra) definition of the quadratic form translates to the square of the action on the spinors,
so that the two vectors in (\ref{oddDirac}) have 
opposite signatures, $(3,1)$ and $(1,3)$ respectively. This implies that, in contrast to the usual Clifford algebra formalism, 
it is not necessary to complexify the Lie
algebra here in order to achieve this change of signature, 
that is necessary to describe the weak interaction. For further details, see Section~\ref{Diracalg}.

To restrict from $\so(6)$ to $\gu(3)$ it is necessary and sufficient to
choose the complex scalar algebra $\gu(1)$, for example 
\begin{align}
\label{u3scalar}
D_{i,il}+D_{j,jl}+D_{k,kl},
\end{align}
which defines a copy of $\su(3)$ with a basis
\begin{align}\label{su3c}
\begin{array}{lccc}
& D_{i,il} - D_{j,jl} & D_{j,jl}-D_{k,kl}\cr
& D_{i,jl}+D_{j,il} & D_{j,kl}+D_{k,jl} & D_{k,il}+D_{i,kl}\cr
& D_{i,j}+D_{il,jl} & D_{j,k}+D_{jl,kl} & D_{k,i}+D_{kl,il}
\end{array}
\end{align}
This copy of $\su(3)$ is chosen for the same reason that governs the choice of $\spl(3,\RR)$ in \cite{MDW},
namely that it lies in a copy of the $G_2$ algebra that acts as automorphisms of the octonions
in the same way on $X$s, $Y$s and $Z$s. This is not just a useful computational tool, but also
has important physical implications in relating spinors to vectors, and therefore relating quantum physics
to classical physics.
The rest of $\so(6)$ then consists of a complex $3$-vector with real components
\begin{align}
\begin{array}{lccc}
& D_{i,jl}-D_{j,il} & D_{j,kl}-D_{k,jl} & D_{k,il}-D_{i,kl}\cr
& D_{i,j}-D_{il,jl} & D_{j,k}-D_{jl,kl} & D_{k,i}-D_{kl,il}
\end{array}
\end{align}

Similarly we restrict from $\so(4)$ to $\gu(2)$ by choosing another scalar, such as
\begin{align}
\label{u2scalar}
X_{lK}\pm D_{I,J}
\end{align}
in which a single change of sign on any of the coordinates $l,I,J,K$ 
reverses the chirality.  
It is not obvious that we have a free choice of sign at this point. 
Later calculations (see Section \ref{GGmodel}) suggest that the negative sign is likely to produce a better model, and that
permutations of $I,J,K$ are associated in some way with permutations of the three generations of elementary fermions.
Let us therefore choose the  scalar $X_{lK}-D_{I,J}$ associated with the $K$ generation. We can then calculate the following basis
for $\su(2)$
\begin{align}
\begin{array}{lccc}
& X_{lK}+D_{I,J} & X_{lI}+D_{J,K} & X_{lJ}+D_{K,I} \cr
\end{array}
\end{align}
The rest of $\so(4)$ then consists of two further \emph{scalar} (not spinor!) representations of $\su(2)$, generated by
\begin{align}
X_{lI}-D_{J,K}, X_{lJ}-D_{K,I}.
\end{align}
Combining $\so(4)$ with $\so(6)$ gives us a group $Spin(6) \circ Spin(4)$ isomorphic to the gauge group
 of
the Pati--Salam model \cite{PatiSalam}. 

\section{The Penrose--Dirac algebra}
\label{Diracalg}
Before we discuss relationships to $SO(10)$ and $SU(5)$ GUTs, it is useful to examine in more detail the relationship between the
Penrose algebra $\so(2,4)$ and the Dirac algebra $\mathbb C\ell(3,1)$, and the corresponding relationship between Penrose twistors and Dirac spinors.
For this purpose we first restrict from $E_8$ to $F_4$, by dropping the labels $i,j,k,l,il,jl,kl$. We then have $16$ real dimensions of spinors/twistors
labelled by $Y$s and $Z$s, together with a symmetry algebra $\so(5,4)$ of dimension $36$ labelled by $D$s and $X$s.
Splitting $\so(5,4)$ as $\so(3)+\so(2,4)$, we see a `generation' symmetry group $Spin(3)$ acting on $I,J,K$, related to the
group $SU(2)_R$ that commutes with $SU(2)_L$ inside $SO(4)$, 
but subtly different from it. Specifically, it is the projection of $SU(2)_R$ onto the $D$ part of the algebra.
Neither of these can be a true generation symmetry group,
as it must mix in some way with mass-momentum symmetries on $L,IL,JL,KL$
to create the mass-changing generation-mixing of elementary particles that is observed experimentally.
See \cite{MDW,Hchirality} for further discussion. 

The remaining $18$ dimensions consist of three $6$-vectors, one for each `generation',
each containing a Lorentz $4$-vector plus two Lorentz scalars which 
between them contain one real parameter and one discrete (binary) parameter, sufficient to encode both mass and some 
appropriate quantum number, probably related to charge.
The spinors/twistors then have the structure of a tensor product of an ordinary $2$-dimensional complex 
spinor for $Spin(3)$ and a $4$-dimensional complex twistor for $Spin(2,4)$. Thus they have an $8$-dimensional complex structure.
But this complex structure is not canonical, as it depends on a choice of complex structure  for the spinor
for $Spin(3)$, from the quaternions $I,J,K$. 
In particular, there is a choice of three different complex structures for three different generations. 
We shall discuss later on how to implement these complex structures inside the Lie group.

 The complex structure on the
twistor, on the other hand, is well-defined (up to complex conjugation), since it is defined by
the action of the scalars in $SU(2,2)$. 
To see this explicitly we need to use the action of the group generators on the spinors,
which is summarised in Appendix A. 
Recall that the complex scalar on a Weyl spinor can be defined by the product of the three
Pauli matrices, here identified with $X_{IL}$, $X_{JL}$ and $X_{KL}$. If we combine this product with $D_L$, then we obtain a square root
of $-1$ that commutes with the whole of $\so(2,4)$. Indeed, this contruction is equivalent to constructing the pseudoscalar
(volume element) of $C\ell(2,4)$. We may therefore define a scalar $\iota$ on the spinors/twistors by
\begin{align}
\begin{array}{l|cccccccc|}
\iota \times & Y_1 & Y_I & Y_J & Y_K & Y_L & Y_{IL} & Y_{JL} & Y_{KL}\cr
= & -Z_{1} & Z_I & Z_J & Z_K & -Z_{L} & -Z_{IL} & -Z_{JL} & -Z_{KL} 
\end{array}
\end{align}
Note that if the subscripts are multiplied by an imaginary unit from $i,j,k,l,il,jl,kl$, then the sign of the complex conjugation operation
in each of the three $X$ operators changes, while the split octonion conjugation is unaffected, so that
\begin{align}
\begin{array}{l|cccccccc|}
\iota \times & Y_l & Y_{lI} & Y_{lJ} & Y_{lK} & Y_{lL} & Y_{lIL} & Y_{lJL} & Y_{lKL}\cr
= & Z_{l} & -Z_{lI} & -Z_{lJ} & -Z_{lK} & Z_{lL} & Z_{lIL} & Z_{lJL} & Z_{lKL} 
\end{array}
\end{align}
In all cases, the distinction between $\iota$ and $-\iota$ is a notation only, and
has no \emph{physical} meaning at this stage. 
To give it a physical meaning, we will need a charge operator to distinguish matter from antimatter,
and a charge conjugation operator to act as complex conjugation
(i.e. to negate $\iota$), so that 
we can express the asymmetry between positive and negative charge
in the real world. 

It is important to note that $\iota$ is not an element of the Lie algebra, only of its enveloping algebra. It is similar to the
element $X_1$, used in \cite{MDW} for defining a complex structure on spinors, being in fact $X_1$ on the labels $U,I,J,K$ and
$-X_1$ on the labels $L,IL,JL,KL$. But we cannot use $X_1$ for a complex structure here, as it does not commute with the two
Lorentz $4$-vectors (\ref{oddDirac}) that are needed as a replacement for the odd part of the Dirac algebra. 
Indeed, $X_1$ acts as the parity symmetry P on the first of these vectors, and as time reversal T on the other.
This means we can use $X_1$ either for the
parity operator $P$, or for the time-reversal operator $T$, as 
will be
discussed in more detail later in this section and in Section~\ref{CPT}.

In order to split the $16$-space of $Y$s and $Z$s first into twistors and then into Majorana/Weyl spinors, we first 
need to find a $2$-dimensional (complex) eigenspace of some element(s) of the algebra, since this is the smallest eigenspace dimension
that is available. It doesn't matter which elements we use, so for 
mathematical simplicity
we use elements with real eigenvalues, say $D_L$ and $X_{KL}$, without regard to their physical meanings. 
All the simultaneous eigenspaces of $D_L$ and $X_{KL}$ are complex $2$-spaces, and any vector in any one of these $2$-spaces lies in a
twistor representation of $\so(2,4)$. There is therefore a $2$-parameter family of twistors
that we could use. Here are complex bases for 
two independent members of this family, each split into eigenspaces of $D_L$ to show the left-handed and right-handed Weyl spinors:
\begin{align}\label{firstpot}
\begin{array}{cc}
(Y_{1+L}+Z_{K+KL},Y_{JL-J}+Z_{I+IL}) & (Y_{1-L}+Z_{K-KL},Y_{J+JL}+Z_{IL-I}) \cr
(Y_{J-JL}+Z_{I+IL},Y_{1+L}-Z_{K+KL}) & (Y_{J+JL}+Z_{I-IL},Y_{L-1}+Z_{K-KL})
\end{array}
\end{align}
The bases for the two rows are compatible, so that coordinate-wise complex linear combinations can be taken in order
to construct all the other twistors in this family.

This construction in some sense identifies Dirac spinors \cite{Dirac} with Penrose twistors,
but highlights an important problem: without the complex structure of the Dirac algebra,
we cannot identify right-handed Weyl spinors with anti-left-handed spinors. Therefore 
these spinors are not yet actual particles, since we need
two Dirac spinors to do what the Standard Model does with one.
We regard this as a feature rather than a bug, since the ability to distinguish four Weyl spinors is an essential
ingredient in a chiral theory.
By the same token, the Penrose algebra $\so(2,4)$ is identified with a `real form' of the Dirac algebra, that is without scalars, and 
(therefore) without complex scalar multiplication. One possible correspondence is, 
up to signs: :
\begin{align}\label{PenroseDirac}
\begin{array}{c|ccc}
& D_{JL,KL} & D_{KL,IL} & D_{IL,JL}\cr
D_{L} & X_{IL} & X_{JL} & X_{KL}\cr\hline
X_L & D_{IL} & D_{JL} & D_{KL}\cr
X_{1} & D_{L,IL} & D_{L,JL} & D_{L,KL}
\end{array}
\qquad
\begin{array}{c|ccc}
&  \gamma_2\gamma_3 & \gamma_3\gamma_1 & \gamma_1\gamma_2\cr
 \gamma_5 & \gamma_0\gamma_1 & \gamma_0\gamma_2 & \gamma_0\gamma_3\cr\hline
 \gamma_0  &\gamma_1\gamma_5 & \gamma_2\gamma_5 & \gamma_3\gamma_5\cr
\gamma_0\gamma_5  &\gamma_1 & \gamma_2 & \gamma_3
 \end{array}
\end{align}
This correspondence ensures that the Lie product matches the Clifford product, except when the former is zero.
Here $X_1$ 
acts as a time-reversal operator.

An alternative is to use $i\gamma_\mu$ instead, so that $X_1$ acts as a parity-reversal operator:
\begin{align}\label{Dirac2}
\begin{array}{c|ccc}
&  \gamma_2\gamma_3 & \gamma_3\gamma_1 & \gamma_1\gamma_2\cr
\gamma_5  & \gamma_0\gamma_1 & \gamma_0\gamma_2 & \gamma_0\gamma_3
\cr\hline
i \gamma_0\gamma_5 & i\gamma_1 &i \gamma_2 & i\gamma_3 \cr
i\gamma_0 &i\gamma_1\gamma_5 & i\gamma_2\gamma_5 & i\gamma_3\gamma_5
\end{array}
\end{align}
These two cases correspond to 
the cases denoted $Pin(3,1)$ and $Pin(1,3)$ respectively in \cite{Pin},
and are therefore physically different models. We do not yet have enough detail to be able to distinguish them
in any physically significant way,
so we leave this question until Section~\ref{CPT}.

First we need to understand how a pair of twistors matches to a pair of Dirac spinors, in order to understand the
difference between right-handed and anti-left-handed Weyl spinors. A useful place to study this problem is the
Georgi--Glashow model, which makes this distinction by distinguishing 
the natural representation of $SU(5)$ from its complex conjugate. However,
it is important to realise that the complex structure of the $SU(5)$ representations has \emph{a priori} nothing
whatever to do with the complex structure defined by $\iota$. It is the relationship between these two notions
of complex structure that we need to elucidate, in order to understand the elementary particles, and especially
to understand why there are three generations of elementary fermions.

\section{The Georgi--Glashow $SU(5)$ model} 
\label{GGmodel}
A feature of the world we live in is that the symmetry between the three generations is clearly broken, since only
the first generation is visible in ordinary matter in ordinary circumstances. 
To investigate further, we must break the symmetry of the three generations by restricting from
$\so(4)$ to $\su(2)_L + \so(2)$. This subalgebra can be combined with the standard model breaking of $\so(6)$ to $\su(3)$
and then embedded in Georgi--Glashow $\su(5)$.
A copy of $\su(5)$ can be defined for any choice of generation, 
by mixing the scalars of $\gu(2)$ and $\gu(3)$ together, for example into
\begin{align}
\label{u5scalar}
X_{lK}-D_{I,J}+ 
D_{i,il}+D_{j,jl}+D_{k,kl}.
\end{align}

The extra $12$ dimensions of $\su(5)$ then form a complex tensor product of a $3$-vector of $\su(3)$ with a $2$-vector of $\su(2)$,
so can be written either as a doublet of complex $3$-vectors (rows) or a triplet of complex $2$-vectors (columns): 
\begin{align}\label{su5gens}
\begin{array}{cccccc}
D_{i,l}+X_{ilK}& D_{j,l}+X_{jlK}& D_{k,l}+X_{klK} \cr 
 D_{il,l}-X_{iK}& D_{jl,l}-X_{jK}& D_{kl,l}-X_{kK}  \cr\hline 
 X_{iI}-X_{ilJ}&X_{jI}-X_{jlJ}&X_{kI}-X_{klJ}\cr
 X_{iJ}+X_{ilI}&X_{jJ}+X_{jlI}&X_{kJ}+X_{klI}
\end{array}
\end{align}
The rest of $\so(10)$, that is the part that is orthogonal (with respect to the Killing form) to all of the subalgebras
$\su(5)$, $\so(6)$ and $\so(4)$ considered so far, can then be obtained by taking the opposite sign combinations in this table.

Now the centralizer of $\su(5)$ in $\elie_{8(-24)}$ is $\su(2,3)$, containing the Penrose algebra $\so(2,4)=\su(2,2)$
together with a scalar and a twistor. 
We already know that the scalar is 
(\ref{u5scalar}), since this commutes with both $\su(5)$ and $\so(2,4)$. We can then calculate the twistor
from any generator, obtained by enforcing the condition that it commutes with a generating set for $\su(5)$.
The calculations are summarised in Appendix A, and give the following result:
\begin{align}\label{su23twistor}
\begin{array}{cccc}
Y_{1-lK}+Z_{K+l} & Y_{K+l}+Z_{1-lK} & Y_{L+lKL}+Z_{KL+lL} & Y_{KL-lL}-Z_{L-lKL}\cr
Y_{I-lJ} + Z_{J-lI} & Y_{J+lI}-Z_{lJ+I} & Y_{IL-lJL}-Z_{JL-lIL} & Y_{JL+lIL} + Z_{IL+lJL}
\end{array}
\end{align}
This twistor is not acted on by the gauge group, so cannot reasonably be said to consist of particles.
But it is acted on by the Lorentz group, so it might be interpreted as a (fermionic) `vacuum'.

The $12$ rotational generators of $\su(2,3)$
are $X_1$ and the double-index $D$s from (\ref{PenroseDirac}), together with
the left half of (\ref{su23twistor}), plus the extra scalar. 
They split into the structure
$\gu(1) + \su(2) + \su(3)$ as follows.
The two scalars split into
\begin{align}
&X_{lK}-D_{I,J}+ 
D_{i,il}+D_{j,jl}+D_{k,kl} - 5X_1,\cr
&X_{lK}-D_{I,J}+ 
D_{i,il}+D_{j,jl}+D_{k,kl}+X_1,
\end{align}
the first of which generates the $\gu(1)$ factor, and the second of which lies in $\su(3)$.
The rest of the generators are allocated as shown:
\begin{align}\label{su123gens}
\begin{array}{lccc}
\su(2): & D_{L,IL}+D_{JL,KL} & D_{L,JL}+D_{KL,IL} & D_{L,KL}+D_{IL,JL}\cr
\su(3): &  D_{L,IL}-D_{JL,KL} & D_{L,JL}-D_{KL,IL} & D_{L,KL}-D_{IL,JL}\cr
 &   & Y_{1-lK}+Z_{K+l} & Y_{K+l}+Z_{1-lK}\cr
 & & Y_{I-lJ}+Z_{J-lI} & Y_{J+lI}-Z_{I+lJ}
\end{array}
\end{align}
This compact algebra is of course isomorphic to the gauge algebra of the standard model, but is quite different from it.

\begin{spec}
At this stage we have two copies of $\gu(1)+\su(2)+\su(3)$, one of which is the compact part of the `spacetime/vacuum' copy of $\su(2,3)$,
while the other is the gauge algebra of the standard model. It is therefore possible to use an isomorphism between the two to
relate properties of the standard model to properties of macroscopic spacetime, and potentially explain (some of) the
unexplained parameters via 
a `symmetry-breaking' of this kind. 

An isomorphism defines 
$12$ arbitrary parameters. 
But the isomorphism does not extend to the non-compact part of $\su(2,3)$, and in particular does not
 extend to the non-compact part of $\so(1,3)$. This would seem to imply that 
at most $12$ of the $24$ standard model parameters could be considered to be universal constants,
while the other $12$ would vary, in particular with the energy scale.
 \end{spec}
 
\section{Representations of $SU(5) \times SU(2,3)$}
\label{GGparticles}
The remaining $200$ dimensions of $E_8$ split into two $50$-dimensional complex irreducible
representations, one of the form $5\otimes 10$, the other of the form $10\otimes 5$, where the first factor
is a representation of $SU(5)$ and the second factor is a representation of $SU(2,3)$.
The Standard Model splits $SU(5)$ into $SU(3)\times SU(2)$, so splits the corresponding $5$ as $3+2$ and $10$ as $1+3+6$.
The Penrose algebra splits the other $5$ as $4+1$ and $10$ as $4+6$, in which the $4$s are both twistor representations.
Combining these splittings gives
real dimensions
\begin{align}
2\times(2+3)\times(4+6) & = 16+24+24+36,\cr
2\times(1+3+6)\times (1+4) & = 2+6+12+8+24+48.
\end{align}
In particular, there are $15$ twistors that split in the same way that the Weyl spinors representing the
elementary fermions split in the
Georgi--Glashow model. 

In order to calculate these explicitly, we first list a suitable coordinate of the $2\times(2+3)$ real $6$-vectors of $\so(2,4)$, say
\begin{align}
D_l, D_I, D_J, D_K, D_i,D_j,D_k,D_{il},D_{jl},D_{kl}
\end{align}
and the $2\times(1+3+6)$ scalars, namely
\begin{align}
\begin{array}{ccc}
X_{lJ}-D_{K,I}& X_{lI}-D_{J,K}\cr\hline
D_{i,jl}-D_{j,il} & D_{j,kl}-D_{k,jl} & D_{k,il}-D_{i,kl}\cr
D_{i,j}-D_{il,jl} & D_{j,k}-D_{jl,kl} & D_{k,i}-D_{kl,il}\cr\hline
D_{i,l}-X_{ilK}& D_{j,l}-X_{jlK}& D_{k,l}-X_{klK} \cr  
 D_{il,l}+X_{iK}& D_{jl,l}+X_{jK}& D_{kl,l}+X_{kK}  \cr
 X_{iI}+X_{ilJ}&X_{jI}+X_{jlJ}&X_{kI}+X_{klJ}\cr
 X_{iJ}-X_{ilI}&X_{jJ}-X_{jlI}&X_{kJ}-X_{klI}
\end{array}
\end{align}
The commutators of these elements with the twistor (\ref{su23twistor}) are then the $15$ twistors we require.
For simplicity let us just calculate the terms involving $Y_1$ or $Y_i$,  
first in the $2+3$ case,
as follows:
\begin{align}
[D_l,Y_{K+l}+Z_{1-lK} ] & = Y_{1-lK}-Z_{K+l}\cr
[D_I,Y_{I-lJ} + Z_{J-lI}] & = Y_{1+lK} +Z_{K-l}\cr
[D_J,Y_{J+lI}-Z_{lJ+I}] & = Y_{1+lK}+Z_{K-l} \cr
[D_K,Y_{K+l}+Z_{1-lK} ] & = Y_{1-lK}-Z_{K+l}\cr
[D_i,Y_{1-lK}+Z_{K+l}] & = -Y_{i-ilK} -Z_{iK-il} 
\end{align}
These twistors represent leptons and anti-down quarks in the Georgi--Glashow scheme.
Similarly the $1+3+6$ case gives us
\begin{align} 
[X_{lI}-D_{J,K},Y_{I-lJ} + Z_{J-lI}]/2 & = -Y_{1+lK} + Z_{K-l}\cr
[D_{j,k}-D_{jl,kl} , Y_{1-lK}+Z_{K+l}]/2 & =-Y_{i+ilK} + Z_{iK+il}\cr
[D_{i,l}-X_{ilK},Y_{K+l}+Z_{1-lK} ]/2 & =Y_{i-ilK} - Z_{iK-il}\cr
[X_{iI}+X_{ilJ},Y_{J+lI}+Z_{lJ+I}]/2 & = -Y_{i+ilK}-Z_{iK+il}
\end{align}

The particles are therefore labelled as follows.
\begin{align}
\label{GGnames}
\begin{array}{cc}
\nu:&-Y_{1-lK}+Z_{K+l}\cr
e: &Y_{1+lK} +Z_{K-l}\cr
\bar e :&  Y_{1+lK} - Z_{K-l}
\end{array}\qquad
\begin{array}{cc}
\bar d: & -Y_{i-ilK} -Z_{iK-il}\cr
\bar u: & -Y_{i+ilK} + Z_{iK+il}\cr
d:&-Y_{i-ilK} + Z_{iK-il}\cr
u:& -Y_{i+ilK}-Z_{iK+il}
\end{array}
\end{align}
Notice, however, that these are twistors, not Weyl spinors as in the Georgi--Glashow model, 
so they contain both left-handed and right-handed spin states. In other words, this model, unlike the Standard Model, makes a distinction
between right-handed and `anti-left-handed' states.
In order to
clarify this distinction, to enable us to describe the reduction to the Standard Model in detail,  
we need a charge operator, which in the Georgi--Glashow model is made out of a combination of
the elements $X_{lK}$ and $D_{I,J}$, together with
\begin{align}
S_l:= D_{i,il}+D_{j,jl}+D_{k,kl}.
\end{align}
Composing these operators with the scalar $\iota$ gives us the following real eigenvalues on the
particle states listed above:
\begin{align}
\begin{array}{c|cccc}
\mbox{Particle} & X_{lK} & D_{I,J} & S_l & S_l/6 + D_{I,J}/2\cr\hline
\nu & 1 & 1 & -3 & 0\cr
e & -1 & -1 & -3 & -1\cr
\bar e & -1 & 1 & 3 & 1\cr
d & -1 & -1 & 1 & -1/3\cr
\bar d & -1 & 1 & -1 & 1/3\cr
u & 1 & 1 & 1 & 2/3\cr
\bar u & 1 & -1 & -1 & -2/3\cr\hline
\end{array}
\end{align}
This table suggests that the operator $S_l/6+D_{I,J}/2$ represents a charge operator for the $K$ generation.
It also indicates that it is unlikely to be possible to construct a `universal' charge operator that commutes with
the generation symmetries in an $E_8$ model of this kind. 

\begin{spec}
Note that in every case except the neutrinos,
the change from particles to antiparticles
(i.e. twistors to anti-twistors) is simply a change from $Z$ to $-Z$. 
The change from neutrinos to anti-neutrinos, however, is a change within a single twistor. 
Hence for consistency it may be better not to
use the term `antineutrino' at all, but rather to distinguish left-handed and right-handed neutrinos.
If we do this, then the `anti' neutrinos, in the sense of
charge conjugation rather than parity reversal, are not particles, but are simply the vacuum.
This is the twistor part of the group $SU(2,3)$, that acts as a kind of `supersymmetry' group on
the spacetime/vacuum.
The reason for calling this
`supersymmetry' is simply that 
the twistor acts by mapping fermions to bosons and bosons to fermions.
But we do not suggest interpreting this supersymmetry as giving rise to new particles.
\end{spec}

We have now allocated $120$ real dimensions of fermions to representations of $SU(5)$
in a way that is compatible with the Georgi--Glashow $SU(5)$ GUT,
but with Weyl spinors (two complex dimensions) extended to twistors (4 complex or 8 real dimensions).
Most importantly, the splitting of fermions into 
twistors and anti-twistors 
is independent of the splitting of each twistor into left-handed and right-handed Weyl spinors.  
For any particular choice of Lorentz group, every twistor splits
into a left-handed and a right-handed Weyl spinor. 

\section{C, P and T symmetries}
\label{CPT}
In order to produce a truly chiral model, it is necessary to implement a relationship
between the `weak handedness' of $\so(4)$, acting on $l,I,J$ and $K$,
and the 
`Lorentz handedness' of $\so(1,1)$, acting on $U$ and $L$.
Assuming that we want this `chirality' to be 
independent of (electron) generation, labelled by $I,J,K$, we must mix $l$ with one of $U,L$. The elements that do this
are 
$X_l$ and $X_{lL}$ respectively, both of which
commute with Lorentz $\so(1,3)$.
By design, they both anti-commute with the scalar $\iota$, which suggests an interpretation as charge conjugation,
possibly combined with one or both of the parity and time-reversal symmetries P and T.

Hence the algebra generated by the $C$, $P$ and $T$ operators is $\so(3,1)$, acting on the labels $1,l,U,L$.
The precise labelling of the individual operators is less important than the consideration of the algebra as a whole.
The discussion in \cite{Pin}, for example, indicates that there is some disagreement about the correct way to
label these operators, and that there may even be a mathematical inconsistency in some of the standard conventions.
It should be noted for example that the only one of the six dimensions of $\so(3,1)$ that is a symmetry of both electromagnetism and the
weak force is $D_L$, 
but that this universal symmetry is  labelled somewhat inconsistently in \cite{Pin}
as both CPT and $\Lambda_P\Lambda_T$. Here we opt for the notation $PT$ for this particular operator,
consistent with our earlier description of P and T. 

One of the main themes of \cite{Pin} is a discussion of the (physical) difference between $Pin(1,3)$ and $Pin(3,1)$,
which is equivalent to the difference between $P=X_1$ and $P=X_L$.
On the basis that the $P$ operator should not affect the mass (located in $L$), but the $T$ operator should act on both
energy and mass,
we assume the former for the ensuing discussion, while recognising that this convention may
not be correct, or consistent with those of \cite{Pin} or any other particular source:
\begin{align}
\begin{array}{ccc|ccc}
C & P & T & PT & CT & CP\cr
D_l & X_1 & X_L & D_L & X_{lL} & X_l
\end{array}
\end{align}
With this convention the even products are Lorentz-invariant, and the odd products are not.
This convention is therefore consistent with the naive interpretation of the C, P and T anti-symmetries of classical
electromagnetism, whereby the negation of any \emph{even} number of directions of charge, space or time is a symmetry of the theory.

The mathematical structure of $\so(3,1)$ suggests re-arranging this table to separate the rotations (not involving $T$) from the
boosts (involving $T$):
\begin{align}
\begin{array}{ccc|ccc}
C & P & CP & PT & CT & T\cr
D_l & X_1 & X_l & D_L & X_{lL} & X_L
\end{array}
\end{align}
which separates massless terms (rotations, not involving $L$) from massive terms (boosts, involving $L$).
But we could also re-arrange to separate  $C$ from the rest:
\begin{align}
\begin{array}{ccc|ccc}
 P & T & PT &C& CT & CP\cr
 X_1 & X_L & D_L &D_l & X_{lL} & X_l
\end{array}
\end{align}
and thereby separate the symmetries of the weak interaction (not involving $l$) from the others. 
Similarly, we could separate $P$:
\begin{align}
\begin{array}{ccc|ccc}
C & CT & T & PT & P & CP\cr
D_l & X_{lL} & X_L & D_L & X_{1} & X_l
\end{array}
\end{align}
although it is not immediately obvious whether this arrangement has any 
physical significance.

Now the `volume element' or (pseudo)scalar for the copy of $\so(3,1)$ generated by $C$, $P$ and $T$
is $D_l\circ D_L$. 
This operator anti-commutes with the volume elements of all four of the following important algebras:
\begin{itemize}
\item the Lorentz or `momentum-energy' $\so(1,3)$ acting on $1,IL,JL,KL$, and with volume element $X_{IL}\circ X_{JL}\circ X_{KL}$;
\item the `mass-momentum' $\so(4)$, acting on $L,IL,JL,KL$, and with volume element $D_{L,IL}\circ D_{JL,KL}$;
\item the weak $\so(4)$, acting on $l,I,J,K$, and with volume element $X_{lK}\circ D_{I,J}$;
\item the twistor $\so(2,4)$, acting on $1,U,L,IL,JL,KL$, and with volume element $D_L\circ X_{IL}\circ X_{JL}\circ X_{KL}$.
\end{itemize}
In other words $D_l\circ D_L$ swaps left-handed and right-handed spinors for all four of these algebras simultaneously,
so provides a well-defined notion of `handedness' that applies to all these different concepts of `spinor'.

\begin{spec}
Whatever the details of this allocation of C, P and T symmetries, the Lie algebra that contains all of them and the Lorentz $\so(1,3)$
is isomorphic to $\so(3,4)$.
This suggests that a more satisfactory model might be built using $\so(3,4)$ instead of $\so(2,4)$. However, it is also clear that this
algebra contains symmetries that do not exist in the real world. Therefore it is necessary to use only a subalgebra, but it must be one
that involves all  $7$ vector coordinates, and therefore it can only be $\glie_{2(2)}$. A model of chirality based on this idea is explored in
\cite{Hchirality}. Its 
most notable attribute is that the Lorentz symmetries then become approximate symmetries only, valid exactly in $2+1$
spacetime dimensions, but only approximately in $3+1$ dimensions, as previously suggested in \cite{Lorentz}.
\end{spec}

 \section{Symmetry-breaking}
 The physical interpretation of the $C$, $P$ and $T$ operators is now significantly more complicated than in the
 Standard Model, since they act not on a single Dirac spinor, consisting of a pair of Weyl spinors, but on a set of four
 distinct Weyl spinors, forming a pair of twistors. In particular, these operators generate a group of order $8$ modulo scalars,
 rather than $4$ as in the Standard Model.
 Therefore certain equivalences that are taken for granted in the Standard Model are not valid in this larger model.
 This is in addition to the confusion already caused \cite{Pin} by identifying the $PT$ operator with the CPT symmetry.
 In particular, we must make a clear distinction between T and CP, between P and CT, and between C and PT.
 
 Most importantly, the operator $C$ maps massive particles to antiparticles, with the opposite charge, while the operator $PT$ does not.
 On the other hand, $PT$ maps neutrinos to anti-neutrinos, while $C$ creates neutrinos and anti-neutrinos out of the 
 quantum vacuum. Hence the standard interpretation of antiparticles as particles moving backwards in time can only be
 obtained by mixing $C=D_l$ with $PT=D_L$ into a single complex number. There is then a parameter that describes this mixing
 as a complex phase, which expresses some aspect of the asymmetry between particles and antiparticles. Normally, this would
 be interpreted as an asymmetry under time-reversal, and therefore under CP, so would be described as a CP-violating phase.
 Similarly, it is necessary to mix $P=X_1$ with $CT=X_{lL}$ and $T=X_L$ with $CP=X_l$. 
 In principle, at least, these three CP-violating phases could all be different.
 
 \begin{spec}
 There are two explicit CP-violating phases in the Standard Model, one in the Cabibbo--Kobayashi--Maskawa (CKM)
matrix \cite{Cabibbo,KM} and one in the Pontecorvo--Maki--Nakagawa--Sakata (PMNS) matrix \cite{Pontecorvo,MNS}. 
We suggest the third angle here is likely to be the electro-weak mixing angle. All three of them can be expressed 
either additively, as above, or multiplicatively, in terms 
of the volume element of $\so(3,1)$, which is:
\begin{align}
D_l\circ D_L = X_l\circ X_L = - X_1 \circ X_{lL}.
\end{align}
 It is therefore possible that the angles in the Standard Model are mixtures of the three angles suggested here.
 For example, it is 
 suggested in \cite{icosa} that the sum of the electro-weak mixing angle 
 and the CP-violating phase
 of the CKM matrix 
 might be more fundamental than the electro-weak mixing angle itself.
 It is also possible, but very hard to test, that the sum of all three of these angles might be exactly $270^\circ$,
 representing a perfect rotation from real to imaginary numbers.
 \end{spec}
The fact that the real world is not invariant under the  C symmetry, here represented by the operator $C=D_l$,
goes back to the discovery of the electron in the late 19th century. On the other hand, invariance under PT, represented by $PT=D_L$,
is an important feature of classical electromagnetism, and also
seems to be a feature of our model, at least in an inertial frame. 
Non-invariance under P, represented here by $X_1$, was shown by the Wu experiment \cite{Wu} in 1957.
On the other hand, invariance under CT, represented here by $X_{lL}$, is claimed in \cite{Pin} to be a necessary condition
for any model, in order to avoid negative energy states. In our model, however, $X_{lL}$ not only swaps massive particles with antiparticles,
but also swaps neutrinos with the vacuum.
It is therefore not technically a symmetry of the model, but instead 
avoids negative energy states 
by extracting energy
from the vacuum. Non-invariance under CP, represented here by $X_l$, was 
 demonstrated in 1964 \cite{CPexp}, but
 it is also possible that
 this should be interpreted here as non-invariance under T, represented by $X_L$.
 
 \section{Non-inertial frames of reference}
 We recall that non-relativistic quantum mechanics is modelled on the group $SU(2)=Spin(3)$ and its Lie algebra $\so(3)$,
 so that it is invariant under all of the $C$, $P$ and $T$ operators. Relativistic quantum mechanics, introduced by Dirac \cite{Dirac},
 extends the group to $SL(2,\CC)=Spin(3,1)$ and the algebra to $\so(3,1)=\so(1,3)$, and reduces the symmetries to the
 even combinations of $C$, $P$ and $T$. This ensures that the model is invariant under inertial changes of reference frame,
 which fix the $PT$ operator. If one wants to extends the group further, in order to gain some non-inertial reference frames,
 then it is necessary to lose the invariance under $PT$. In our model, the largest such group that commutes with the gauge group
 and preserves the fundamental distinction between fermions and bosons is $SU(2,2)=Spin(2,4)$.
 
 As is well-known, this group plays a fundamental role in the approach to non-inertial motion and gravity taken by Penrose \cite{RR},
 which has recently been revived by Woit \cite{WoitSO24}. In our model, this group also commutes with the charge operator, and
 therefore preserves the fundamental distinction between particles and antiparticles. It is apparently not big enough to
 contain General Relativity, since it does not contain the general covariance group $GL(4,\RR)$, but may nonetheless provide
 a useful extension of Special Relativity to include observers in non-inertial rest frames.
 
What is important here is that $\so(2,4)$ commutes with the gauge groups of the standard model, and with
the charge operator, and with the generation and colour symmetries, and preserves all the particle types. It is
therefore capable of extending the Standard Model to non-inertial frames, in a non-trivial way. This is a necessary
stepping-stone on the way to implementing a quantum theory of gravity that is compatible with the Standard Model.
But first, note that the ability to deal with non-inertial motion is particularly important for the analysis of experiments
performed on a rotating Earth. The model may therefore provide new interpretations of old experiments,
since the standard interpretations do not take into account the non-inertial motion of the experiment.

From this point of view, there is a choice of ways to interpret the Wu experiment  
that observed a correlation between the directions of atomic spin and momentum of the ejected electron.
Although this effect is usually called parity-violation, the Lorentz-invariant name for this effect in our model is $CT$-violation.
In the Standard Model, $CT$-violation is not permitted, since it implies the existence of negative energy states.
But if we add in to the energy calculation the energy of the non-inertial motion of the experiment, then this non-inertial
motion itself cancels out the negative energy that would otherwise be in the theory. Hence there is no physical contradiction here.
This interpretation also explains why the effect cannot be detected unless all other sources of energy are rigorously excluded
from the experiment. Moreover, the energy transfer required to eliminate the negative energy state can only come from
a quantum gravitational interaction, since all other effects have been eliminated.

An alternative way to interpret this result is as $P$-violation of electromagnetism, rather than as $CT$-violation of the
weak interaction. The two interpretations are equivalent, via the general principle of relativity, but one explanation
is more appropriate in an inertial frame, and the other in the (non-inertial) laboratory frame.
Specifically, the $P$ operator
$X_1$ does not commute with $SO(1,3)$, and therefore it represents a change between an inertial frame of reference and
a non-inertial frame. Therefore parity violation in this model is a direct consequence of switching between an inertial frame and a
non-inertial frame in which to describe the \emph{electromagnetic} interaction between the electron and the proton/nucleus involved in beta decay. 
It is clear that the laboratory
frame of reference in which the experiment was conducted and analysed was not an inertial frame.
It is perhaps unfortunate that this obvious fact is routinely ignored.

Of course, a simple rotating frame of reference is unlikely to be sufficient for this explanation, since conservation of angular
momentum and conservation of spin would normally be expected to rule out any transfer of energy. However, a complicated motion involving
two or more rotations, such as any laboratory on Earth experiences, is a different matter entirely. There are many parameters that might
in principle affect laboratory measurements, although experiment suggests that most of them do not. For example, there is little or no evidence that 
experiments are materially affected by the latitude of the experiment, the time of day, the time of year, or the phase of the Moon, or 
even on
more esoteric parameters such as the tilt of the Earth's axis or the inclination of the orbit of the Moon. 
Yet there are some anomalies, such as the muon $g-2$ anomaly 
\cite{muong-2,muontheory,muonHVP,Fodor} or the $W$ mass anomaly \cite{WZ,WZtoy}, that might in principle
demonstrate such a (quantum gravitational) effect.

 Most obvious are the long-range neutrino oscillation experiments
 \cite{oscillation,neutrinos,SNO}, in which the copies of
the Lorentz group used in the different laboratories are substantially different. 
The simplest single parameter to measure this change in Lorentz group is the change in the direction of
the gravitational field, although other parameters may also have (presumably smaller) effects.
Even in a very small experiment
of diameter 5 meters, the direction of the gravitational field changes by almost $10^{-6}$ radians across the experiment, 
which  
should be taken into
account in the analysis of precision measurements of, say, 
the muon gyromagnetic ratio. The most likely reason why this particular experiment is susceptible to a correction of this kind
is that the CP symmetry of electromagnetism only holds relative to a fixed copy of the Lorentz group,
so that CP-violations must be taken into account 
as the Lorentz group changes across 
the experiment. 

\begin{spec}
Similarly, the apparent `CP-violation' of neutral kaon decay might also be interpreted as the result of using the same copy of the
Lorentz group across the 57 feet of the experiment. If we correct for the change in direction of the gravitational field over this distance,
we obtain a quantitative postdiction of this CP-violation that agrees with the experiment. Moreover, since the change in gravitational field
manifests itself as a change in a kaon eigenstate, it permits an interpretation of quantum gravity in terms of `virtual kaons',
as suggested by Hossenfelder \cite{hossenfelder}. An alternative interpretation  
is to suppose that the change of kaon eigenstate is accompanied by,
or even caused by, an interaction
with a low energy neutrino,  
which would implicate the neutrinos in interactions that convey gravitational
information \cite{verlinde} between elementary particles. 
\end{spec}

\begin{spec}
 The essential point is that $\so(2,4)$ contains an $8$-parameter family of subalgebras $\so(1,3)$, and can therefore
 express coordinate transformations between reference frames that are not inertial with respect to each other.
  It should be noted that 
the $8$-parameter family of copies of $\so(1,3)$ in $\so(2,4)$ is not related to the obvious $9$-parameter family of
copies of $\so(1,3)$ in $\spl(4,\RR)$, so that there is no direct connection between these non-inertial motions and
the general theory of relativity. However, there \emph{is} a close connection instead to the $8$-parameter family of
copies of $SL(2,\CC)$ in $SL(4,\RR)$, and therefore to a \emph{different} family of subalgebras isomorphic to
$\spl(2,\CC)=\so(1,3)$ in $\spl(4,\RR)$. Whether this other family provides a viable alternative to general relativity
remains to be seen, but is beyond the scope of this paper. 
\end{spec}

\section{Mass and the Dirac equation}
The momentum and energy terms in the Dirac equation anti-commute with $\gamma_5$, that is with $D_L$. The latter represents
the so-called CPT symmetry (here denoted $PT$ for reasons already explained) of electro-weak interactions. Therefore the mass term in
any CPT-invariant Dirac equation must also anti-commute with $D_L$, and commute with the Lorentz group. We therefore get 
ten linearly independent `mass planes', whose real
and imaginary parts are as follows:
\begin{align}
\begin{array}{c|ccc|cccccc}
X_{lL} & D_{I,L} & D_{J,L} & D_{K,L} & X_{iL} & X_{jL} & X_{kL} & X_{ilL} & X_{jlL} & X_{klL}\cr
X_l & D_I & D_J & D_K & X_i & X_j & X_k & X_{il} & X_{jl} & X_{kl}
\end{array}
\end{align}
In each case, the top element is a boost, and the bottom one a rotation. 

Various slightly different implementations of the Dirac equation are conceivable, along the lines of
\begin{align}
\label{Diraceqn1}
 \hbar (X_1\partial_{ct} + D_{IL}\partial_x +D_{JL}\partial_y+D_{KL}\partial_z)\psi  \pm m_ec D_{I,L}\psi = 0,
\end{align}
where $m_e$ is the electron mass, corresponding to the $I$ generation label. Here $D_{I,L}$ squares to $+1$, and commutes with
all the other coefficients $X_1,D_{IL},D_{JL},D_{KL}$, so that this is a textbook example of a Dirac equation. 
Conversion to the momentum-space equation is a little different from usual, since the non-associativity of the octonions affects the calculations.
We need in effect to change the mass term $D_{I,L}$, that squares to $+1$, to a term that squares to $-1$, but the `obvious' change to
$D_I$ does not work, because $D_I$ does not commute with the energy-momentum terms. What does seem to work is to change
$L$ to $l$, while leaving all the other labels unchanged.

We find in fact that the electrons in the three generations satisfy the following versions of the Dirac equation:
\begin{align}
\label{Diraceqn2}
(p_0X_1 + p_1D_{IL} +p_2D_{JL}+p_3D_{KL} - m_ec X_{lI})\psi = 0,\cr
(p_0X_1 + p_1D_{IL} +p_2D_{JL}+p_3D_{KL} - m_\mu c X_{lJ})\psi = 0,\cr
(p_0X_1 + p_1D_{IL} +p_2D_{JL}+p_3D_{KL} - m_\tau c X_{lK})\psi = 0.
\end{align} 
Interestingly, the anti-electrons satisfy the \emph{same} equations, rather than changing the sign of the
mass (or energy) term. 
This is an encouraging sign, as it avoids the necessity of `interpreting' negative energy solutions of the Dirac equation
as anti-particles: there are no negative energy solutions to this equation, and anti-particles have positive mass.
Thus we are led to suggest ten masses associated to the following list of images of the particle names under the action of $X_l$:
\begin{align}
\begin{array}{c|ccc|cccccc}
X_l & D_I & D_J & D_K & X_i & X_j & X_k & X_{il} & X_{jl} & X_{kl}\cr
0 & X_{lI} & X_{lJ} & X_{lK} & D_{i,l} & D_{j,l} & D_{k,l} & D_{il,l} & D_{jl,l} & D_{kl,l}
\end{array}
\end{align}
Of particular note here is the fact that the singlet mass term $X_{lL}$ has disappeared completely, leaving us with only a zero Dirac mass for
neutrinos. In particular, we have now reduced to a list of ten 
{elementary} fermions, with generation labels for electrons and quarks,
but not for neutrinos. This gives us a good first approximation to the Standard Model, but leaves the problem of 
describing neutrino
mass and flavour eigenstates to be dealt with somewhere else in the model.

In particular, there is no Dirac mass available for neutrinos in this model. 
This does not, of course, preclude the possibility that neutrinos have Majorana masses.
Indeed, it is possible to divide the $14$ theoretically available mass planes into the ten 
listed above,
and four more, 
with corresponding mass planes
lying inside $\so(2,4)$:
\begin{align}
\begin{array}{c|ccc}
X_L &D_{IL} & D_{JL} & D_{KL}\cr
X_1 & D_{L,IL} & D_{L,JL} & D_{L,KL}
\end{array}
\end{align}
Here we see a remarkable similarity between the neutrino and electron mass planes, and indeed it is 
really only the non-associativity of the octonions
that distinguishes the neutrinos from the electrons in this regard. This suggests that this non-associativity may be the crucial ingredient in the
definition of mass.
At this stage we have $14$ potential fermion mass terms, two of which, namely $X_{lL}$ and $X_L$, are generation and colour singlets.
It may make sense to interpret these as proton and neutron masses respectively, so that the primary division is into
$10$ charged and $4$ neutral fermions.

\begin{spec}
If we look closely at the proposed neutrino mass planes, 
we find that the first three of these terms have already been allocated to momentum,
while the last three convert between mass in $L$ and momentum in $IL,JL,KL$.
It is therefore not possible to consider
neutrino mass independently of the direction of momentum. Of course, there is no reason to suppose that there is any
correlation between the mass eigenstates and the momentum direction. But there is a reason to suppose 
that the total number of degrees of freedom for neutrinos is less than in the Standard Model. Therefore one should not expect to be
able to specify mass eigenstates, flavour eigenstates, and momentum direction all independently of each other. In particular, the
momentum direction (perhaps relative to the gravitational field) 
may express a relationship between mass eigenstates and flavour eigenstates, and therefore
have an influence on the phenomenon of neutrino oscillations.
\end{spec}

A further important remark is that the proposed mass terms for the quarks do not commute with the charge operator.
This implies that the mass eigenstates are incompatible with the charge eigenstates. This experimentally verified phenomenon
is incorporated into the Standard Model `by hand', using the CKM matrix. In our proposed model, the six mass terms can be
interpreted as a complex $3$-space, with complex structure defined by $l$, and basis $i,j,k$. Hence it is plausible that
a correspondence between the CKM matrix and our model can be constructed. 
This is beyond the scope of the present paper, however.

 \section{Conclusion}
 The attempt to use $E_8$ as the basis for a grand unified theory or a `theory of everything' goes back to the 1980s, but 
 achieved a certain notoriety
 after the publication of Lisi's claimed `theory of everything' \cite{Lisi} in 2007, and Distler and Garibaldi's claimed refutation thereof
 \cite{DG}
 in 2009. The Banff workshop in 2010, which I 
 participated in, was
 organised to try to resolve the issue, but did not succeed in doing so, and left both camps entrenched
 in their positions.
In this paper we have shown that one crucial difference between the two sides can be localised to the 
mathematical definition of a chiral theory, and that a more physically natural
definition of chirality than in \cite{DG} 
has the potential 
to resolve the difficulty. 

We have thus presented a candidate definition that avoids  the Distler--Garibaldi refutation, and permits not only three generations of
fundamental fermions, but also a quantum theory of gravity using Penrose twistors, as recently suggested by Woit \cite{WoitSO24}.  
 It would be an exaggeration to claim that this is a `theory of everything', but it rather strongly suggests that such a theory does in fact 
 exist in $E_8$.
What the $E_8$/twistor model offers, that the Standard Model does not, is a framework in which
to devise methods of calculating corrections for non-inertial motion.
It is suggested that such corrections may be necessary, and possibly sufficient, to resolve certain anomalies involving particles
outside the first generation, such as
the muon $g-2$ anomaly or the $W$ mass anomaly.

\label{discrete}

There is also an important question as to whether generation symmetry should be modelled by a Lie group or a finite group.
It is clear that in the weak interaction, the generations are discrete, so are best modelled by a finite group. 
If we take any copy of $SU(2)$ 
as a continuous generation symmetry group,
and re-interpret the Lie algebra generators ($i$ times the Pauli matrices) as generators for a finite group, they
generate a copy of the quaternion group $Q_8$, which has an automorphism of order $3$ permuting the three
generations. 

Since the weak (flavour) eigenstates and strong (mass) eigenstates are different,
we 
seem to need a finite subgroup of $SU(2)$ that contains at least two
copies of $Q_8$, each with an automorphism of order $3$. It is well-known that 
there is only one such group, namely the binary icosahedral group.
It was suggested in 
\cite{icosa,ZYM} that the appropriate linearised model for quantum field theory in this context is the real group
algebra of this group. This algebra is associative of dimension $120$. The complex group algebra therefore has real
dimension $240$, and is related to $E_8$ in a rather curious and far from obvious way, generally known
as the McKay correspondence \cite{McKay}.
Details are given in Appendix B.

One consequence of this correspondence 
is that the group algebra effectively omits the extra twistor (the `right-handed neutrino')
that we 
interpreted as the vacuum,
and gives us only the $120$ dimensions of fermions that are observed experimentally. Another consequence is that the
algebra is associative, so that we could hope to implement the standard machinery of Clifford algebras more easily than 
in a Lie algebra. A third consequence is that the complexification becomes irrelevant to the underlying structure
of the algebra, so that we could hope to simplify the model to a real algebra of half the dimension.
A fourth consequence is that this simplification mixes the adjoint representations of $SU(5)$ and $SU(2,3)$, and
therefore implies a certain amount of symmetry-breaking. These ideas are explored further in \cite{icosa,ZYM}, which
suggest that the $E_8$ model constructed in this paper may be only an approximation to an underlying discrete reality.

 \section*{Appendix A: details of calculations in $E_8$}
We use the rule
\begin{align}
[D_{a,b},D_{b,c}] = 2b^2D_{a,c}
\end{align}
to check that
the generators for $\su(3)$ in (\ref{su3c}) commute with the scalar:
\begin{align}
[D_{i,jl}+D_{j,il},D_{i,il}+D_{j,jl}]/2 & = D_{jl,il} + D_{j,i} + D_{i,j} + D_{il,jl} = 0\cr
[D_{i,j}+D_{il,jl},D_{i,il}+D_{j,jl}]/2 & = D_{j,il} - D_{jl,i} - D_{i,jl} + D_{il,j} = 0
\end{align}

Next we use the similar rules
\begin{align}
[D_{a,b},X_{bA}]  & = 2b^2 X_{aA}\cr
[X_{aA},X_{bA}] & = 2A\bar A D_{a,b}
\end{align}
to check that the generators for $\su(5)$ in (\ref{su5gens}) commute with the scalar (\ref{u5scalar}):
\begin{align}
[D_{i,l}+X_{ilK}, D_{i,il}+X_{lK}]/2 & = D_{l,il}+D_{il,l} + l^2X_{iK}-(il)^2X_{iK} = 0\cr
[D_{il,l}-X_{iK}, D_{i,il}+X_{lK}]/2 & = D_{i,l} - D_{i,l} - i^2X_{ilK} + l^2X_{ilK} = 0\cr
[D_{i,il}-D_{I,J},X_{iI}-X_{ilJ}]/2 & = -i^2X_{ilI}+I^2X_{iJ}-(il)^2X_{iJ} + J^2X_{ilI} = 0\cr
[D_{i,il}-D_{I,J},X_{iJ}+X_{ilI}]/2 & = -i^2X_{ilJ}-J^2X_{iI}+(il)^2X_{iI} - I^2X_{ilJ} = 0.
\end{align}

For acting on spinors we need the rules
\begin{align}
[D_{a,b},Y_c] & = -Y_{b(ac)}\cr
[D_{a,b},Z_c] & = -Z_{(ca)b}\cr
[X_{aA},Y_{bB}] & = -Z_{\overline{ab}.\overline{AB}}\cr
[X_{aA},Z_{bB}] & = Y_{\overline{ba}.\overline{BA}}
\end{align}
which together enable us to check that the first element of (\ref{su23twistor}) commutes with a set of generators for 
$\su(5)$.  If we first check
\begin{align}
[X_{iI+ilJ}, Y_{1+lK}+Z_{K-l}] & = -Z_{iI+ilJ-ilJ-iI} + Y_{-iJ+ilI-ilI+iJ} = 0,
\end{align}
then the same is true with $i$ replaced by $j$ or $k$. These together generate a subalgebra $\so(3)$ of $\su(3)$,
so we need to check one more generator for $\su(3)$, say $D_{i,il}-D_{j,jl}$. This calculation is easy, because all terms are
associative, and it is almost obvious that the condition holds in this case.
Finally we need to check generators for $\su(2)_L$:
\begin{align}
[X_{lI}-D_{J,K},Y_{1+lK}+Z_{K-l}] & = -Z_{lI-J}+Y_{-I+lJ}+Y_{-lJ+I}-Z_{J-lI}=0\cr
[X_{lJ}+D_{I,K}, Y_{1+lK}+Z_{K-l}] & = -Z_{lJ+I} + Y_{-lI-J} + Y_{J+lI} + Z_{I+lJ} = 0.
\end{align}

\section*{Appendix B: embedding of the binary icosahedral group in $E_8$}

The binary icosahedral group $2I$ has two inequivalent representations in $SU(2)$, related by changing the sign of $\sqrt5$. Let us denote these
$\rep2a$ and $\rep2b$, and denote their symmetric squares $\rep3a$ and $\rep3b$ respectively. 
Full details of the representation theory of $2I$ can be found in \cite{icosa,ZYM}.
Here I describe a particular choice of a copy of $2I$ inside $E_8$, and its representation on the Lie algebra.
We can map $2I$ onto
a subgroup of $SU(2)\times SU(3)$ inside $SU(5)$, 
with representation $\rep2a+\rep3a$ say, and also onto a subgroup of
$SU(2)\times SU(3)$ inside $SU(2,3)$, with representation $\rep2b+\rep3b$. 

If we now calculate the representation of this copy
of $2I$ on the Lie algebra for $E_8$, we first find, in the notation of \cite{icosa}, the adjoint representations of $SU(5)$ and $SU(2,3)$ as
\begin{align}
(\rep2a + \rep3a)\otimes (\rep2a+\rep3a) -\rep1 & = \rep3a + \rep1+\rep3a+\rep5 + \rep2a+\rep4b+\rep2a+\rep4b\cr
(\rep2b + \rep3b)\otimes (\rep2b+\rep3b) -\rep1 & = \rep3b + \rep1+\rep3b+\rep5 + \rep2b+\rep4b+\rep2b+\rep4b
\end{align}
so that the sum of the two consists of two real copies, or one complex copy,  of
\begin{align}
\rep1+\rep3a+\rep3b+\rep5+\rep2a+\rep2b+\rep4b+\rep4b.
\end{align}
Then the $10$-dimensional representations have the form
\begin{align}
\Lambda^2(\rep2a+\rep3a) & = \rep1 + \rep3a + \rep2a +\rep4b\cr
\Lambda^2(\rep2b+\rep3b) & = \rep1 + \rep3b + \rep2b +\rep4b
\end{align}
so that $5\otimes 10$ has the form
\begin{align}
(\rep2a+\rep3a) \otimes (\rep1+\rep3b+\rep2b+\rep4b) & = \rep2a^2 + \rep3a^2 + \rep6^3+\rep4a^2+\rep5^2 +\rep4b\cr
(\rep2b+\rep3b) \otimes (\rep1+\rep3a+\rep2a+\rep4b) & = \rep2b^2 + \rep3b^2 + \rep6^3+\rep4a^2+\rep5^2 +\rep4b
\end{align}
where superscripts denote multiplicities.
We therefore obtain the full representation on $E_8$ as a complex $124$-dimensional
representation of the form
\begin{align}
\rep1 + 3\times(\rep3a+\rep3b)+4\times \rep4a+5\times\rep5 + 3\times(\rep2a+\rep2b) + 4\times\rep4b+6\times\rep6.
\end{align}
This is equivalent to the representation on the complex group algebra, plus an extra copy of
the complex $4$-space $\rep2a+\rep2b$. It is possible to identify the latter with a Cartan subalgbra, so that the
root spaces span the group algebra.

\section*{Acknowledgements}
I thank Tevian Dray and Corinne Manogue for long discussions over many years concerning the mathematics and the physics
of $E_8$, and in particular for detailed comments on the technical differences between the model discussed in this paper
and the similar but subtly different model in \cite{MDW}. I thank Mitchell Porter for many useful comments, and drawing my
attention to \cite{Pin}.

\end{document}